\documentclass[%
 aip,
 amsmath,amssymb,
 preprint,%
]{revtex4-1}

\usepackage{graphicx}
\usepackage{dcolumn}
\usepackage{bm}

\usepackage[utf8]{inputenc}
\usepackage[T1]{fontenc}
\usepackage{mathptmx}
 \usepackage{relsize}

\begin{document}

\preprint{AIP/123-QED}

\title[A finite element discrete Boltzmann...]{A finite element discrete Boltzmann method for high Knudsen number flows}

\author{Karthik Ganeshan}
\altaffiliation{
Corresponding author. Electronic mail: kug46@psu.edu
}
 
\author{David M. Williams}%
\email{dmw72@psu.edu}
\affiliation{ 
Department of Mechanical Engineering, Pennsylvania State University, University Park, Pennsylvania 16802, USA.
}%

\date{\today}

\begin{abstract}
Simulations of the discrete Boltzmann Bhatnagar-Gross-Krook (BGK) equation are an important tool for understanding fluid dynamics in non-continuum regimes. Here, we introduce a discontinuous Galerkin finite element method (DG-FEM) for spatial discretization of the discrete Boltzmann equation for isothermal flows with Knudsen numbers ($\text{Kn}\sim \mathcal{O}(1)$). In conjunction with a high-order Runge-Kutta time marching scheme, this method is capable of achieving high-order accuracy in both space and time, while maintaining a compact stencil. We validate the spatial order of accuracy of the scheme on a two-dimensional Couette flow with $\text{Kn} = 1$ and the D2Q16 velocity discretization. We then apply the scheme to lid-driven micro-cavity flow at $\text{Kn} = 1, 2, \text{and } 8$, and we compare the ability of Gauss-Hermite (GH) and Newton-Cotes (NC) velocity sets to capture the high non-linearity of the flow-field. While GH quadrature provides higher integration strength with fewer points, the NC quadrature has more uniformly distributed nodes with weights greater than machine-zero, helping to avoid the so-called ray-effect. Broadly speaking, we anticipate that the insights from this work will help facilitate the efficient implementation and application of high-order numerical methods for complex high Knudsen number flows.

\end{abstract}

\maketitle

%


\section{\label{sec:intro}Introduction }
Nano- to meso-scale simulations have seen abundant applications, especially in recent years, for porous media flows such as shale gas transport, groundwater flows, fuel cells, and metal-air batteries.~\cite{Wang2016a} These applications typically involve multi-phase, multi-species, and reactive transport with the reactions commonly occurring only at the fluid/solid-boundary interface. While in macroscopic flows, the solutions obtained from the classical Navier-Stokes equations are accurate, there is a degree of departure from their predictions in microscopic flows. This departure is characterized by the Knudsen number ($\text{Kn} = \lambda / L $, where $\lambda$ is the mean free path of the fluid and $L$ is the characteristic dimension), which categorizes a flow as: (a) Continuum flow -- where Kn is assumed to be vanishingly small, typically $\text{Kn} \lesssim 10^{-3}$, (b) Slip flow ($10^{-3} \lesssim \text{Kn} \lesssim 0.1$), (c) Transition flow ($0.1 \lesssim \text{Kn} \lesssim 10$), and (d) Ballistic or free-molecular flow ($\text{Kn} \gtrsim 10$). 
 In the slip flow regime, the departure from the continuum flow occurs only at the boundaries where the no-slip boundary condition is applied. Hence, this can be corrected using a slip boundary condition which allows for a corresponding velocity to be applied based on a theory such as the Maxwell-, Navier-, or higher-order slip  models.~\cite{Wu2008sl,Fukui1990,Mitsuya1993,Hsia1983} However, this correction is not applicable for high Kn flows (i.e., in the transition-flow regime and beyond) as the non-linearity is observed in the bulk-flow and is not isolated to the boundary. This limits the use of conventional CFD, even with corrected boundary conditions in the applications mentioned above.

As a solution to this problem, the Lattice Boltzmann Method (LBM) has gained popularity because of its roots in the Boltzmann Transport Equation (BTE), although it was historically derived from the (failed) Lattice Gas Automata (LGA).~\cite{succi2001lattice} While there was some initial argument that LBM is limited only to continuum flows, and that any noticeable departure is an artifact of discretization, it has been proven that the LBM matches the analytical solutions of the BTE, molecular dynamics simulations, and Direct Simulation Monte-Carlo methods.~\cite{Ansumali2007,Kim2008} The LBM framework constitutes the splitting of the Boltzmann equation into two steps -- collision (which is local to each node), and streaming (which involves moving the value corresponding to a node to its neighbor along the direction of propagation). The simplicity of the framework makes the implementation of this method straightforward while keeping the computation costs low. Traditional LBM has a few disadvantages as the grid is constrained to follow the velocity set used to discretize the momentum-phase of the Boltzmann equation, and the grid spacing is tied to the time step as the streaming step involves a direct translation of values from one node to its neighbors. This induces uniform grids which may not conform to the local geometry. In addition, the framework requires a CFL number of unity, by construction. Furthermore, the coupling between the momentum and position space, as imposed by the LBM discretization, is not mandatory in capturing the correct flow-field dynamics.~\cite{abe1997d,He1997a,Cao1997f}

To obtain a grid-independent variant of the LBM, several methods have been proposed that involve applying the Eulerian framework to the Discrete Boltzmann Equation (DBE). These approaches decouple the momentum phase (the discrete velocities) and the position space (spatial grid) using Finite Difference (FD), Finite Volume (FV), and Finite Element (FE) methods for spatial discretization. 

The FD discretization of the DBE acts as a generalization of LBM but maintains the decoupling of the position and momentum space, allowing the scheme to utilize a non-unity CFL number.\cite{Chen1998} As with most FD schemes, the accurate treatment of curved boundaries is complicated in FD-LBM schemes, and boundary conditions may introduce errors in the evaluation of slip-velocity and mass conservation. High-order FD schemes have been developed for Cartesian and curvilinear grids, such as the 4th-order scheme developed by Hejranfar et al.~\cite{Hejranfar2014a,Hejranfar2014b} However, these schemes typically involve the use of filters to ensure numerical stability, leading to the convergence depending on the filtering coefficients.

The FV formulation of the DBE was first proposed by Amati et al.,~\cite{Amati1997} where the volume-averaged values of the PDFs are obtained from a piecewise linear interpolation on a nonuniform coarse grid. Further developments of FV-LBM have been presented that improve the approximation of conversation laws and allow different element shapes to enable geometric flexibility at the boundaries.~\cite{peng1998lattice,peng1999finite,xi1999finite} Recently, Chen et al. have developed implementations of FV-LBM with a cell-centered approach that improves the accuracy of obtained flow features in complex flows and along curved boundaries.~\cite{Chen2018,Chen2019,Chen2015} 
In some cases, stability of high-order FV-solvers requires the use of Essentially Non Oscillatory ENO~\cite{Harten1986some} and Weighted Essentially Non Oscillatory WENO~\cite{balsara2000monotonicity} schemes. However, the high-order FV formulations require large stencils. These stencils link elements from disparate parts of the mesh, hindering the geometric flexibility of the schemes near the boundary due to the necessity of flux reconstruction. This may also lead to a reduction in the order of accuracy of the formulation near boundaries. 

FE schemes can be considered as a promising alternative to high-order FV schemes, as they possess a more compact stencil. A variety of FE schemes have been proposed to obtain solutions to the DBE. Krivovichev~\cite{Krivovichev2014} and Jo et al.~\cite{jo2009finite} demonstrated the use of Continuous Galerkin FEM (CG-FEM) with equilibrium boundary conditions for viscous flows. However, due to the non-self-adjoint nature of the BTE, these schemes are often susceptible to spurious oscillations.~\cite{jiang1998least} To resolve this problem, Lee et al.~\cite{Lee2001} proposed a characteristic Galerkin approach involving a second-order accurate predictor-corrector step, whereas Li et al.~\cite{li2004least} suggested the use of a least squares scheme with fourth-order accuracy in space and second-order accuracy in time. Discontinuous Galerkin (DG) solvers are generally preferred for discretizing the DBE as they allow for easier parallelization (due to their element-by-element computation) and are better-suited for advection-dominated equations such as the DBE. Shi et al.~\cite{Shi2003} and D\"uster et al.~\cite{Duster2006} have shown that the DG formulation is an efficient solver in obtaining high-order numerical solutions to the DBE. While only a first-order, forward Euler time stepping method was utilized by these researchers, the scheme allows coupling to higher-order time integrators. To further improve the computational efficiency, Min et al.~\cite{Min2011} proposed the decoupling of the DBE into collision and streaming steps, similar to the LBM. Here, the DBE is integrated first using trapezoidal rule and following a transformation of the distribution function, the solution is obtained in two stages with a local collision step and a streaming step, the latter which is treated as an advection-only equation solved using DG-FEM. The Eulerian treatment of the streaming step enables the scheme to be grid-independent. This enables a trivially-diagonalizable mass matrix facilitating efficient computation even with low relaxation times at high CFL numbers. This approach has been used frequently in studying flows through/past cylinders and porous media.~\cite{Wardle2013,Wu2018A,Zadehgol2014} Although high-order temporal integrators are typically used for solving the streaming step, the scheme retains the native second-order accuracy in time due to the use of the trapezoidal rule, or equivalently, a second-order time accurate expansion using Strang splitting. Therefore, for higher temporal accuracy, the space-time coupled DBE is required to be solved with the corresponding higher-order time-integrator.~\cite{Shao2018} Recently, other modifications to DG-DBE schemes have been proposed in order to improve the numerical efficiency.~\cite{Coulette2018,Karakus2019} 

The aforementioned FD, FV, and FE schemes share the ability to operate with non-unity CFL numbers, and (for the FV and FE schemes) to operate on unstructured grids. However, these schemes have only been applied to continuum flows in conjunction with small sets of discrete velocity directions. To capture non-continuum effects, Jaiswal et al.~have developed DG-based solvers for the BTE with Fourier-transform-based discretizations of velocity space capable of handling the full Boltzmann collision operator.~\cite{jaiswal2019discontinuous} Of course, this comes at a significant computational cost due to the complexity of the associated integrals. To decrease the cost of these schemes, Guo et al.~\cite{Guo2013} developed a FV Discrete Unified Gas Kinetic Scheme, using the BGK-collision operator and velocity discretizations similar to the LBM. Theoretically speaking, this scheme is capable of generating accurate solutions at all Knudsen numbers. However, with higher Knudsen numbers and non-linearity in the flow field, a larger velocity set is still required, increasing the computational cost. In addition, although there is no decoupling of the collision and streaming steps, the use of trapezoidal rule in time integration limits the scheme to second-order accuracy in time. 

In this manuscript, we present a fully implicit DG-DBE method implemented for high-Kn flows with high-order accuracy in both time and space. The paper is structured as follows. In section \ref{sec:methods}, we present the foundation of the DBE and the corresponding discretization using DG-FEM, along with the associated velocity sets and boundary conditions. In section \ref{sec:results}, we show the high-order accuracy of the scheme for high-Kn Couette flow, and we examine its performance in conjunction with various velocity sets. Thereafter, we apply the scheme to a highly non-linear high-Kn lid-driven micro-cavity flow, to showcase the flexibility of the method for various flow regimes. Finally, some concluding remarks are provided in section~\ref{sec:conclude}.

\section{\label{sec:methods}Methods }
\subsection{The Discrete Boltzmann Equation}
\subsubsection{The BGK-Boltzmann equation}
Let us begin with the isothermal body-force-free BTE %
\begin{equation}
\label{eqn:BTE}
    \frac{\partial f}{\partial t} + \boldsymbol{\xi}\cdot\nabla f = \boldsymbol{\Omega} \equiv -\frac{1}{\tau}\left(f - f^{eq} \right),
\end{equation}
which is a 6+1-dimensional equation with three dimensions each in space and velocity, and one dimension in time, forming the full phase space for the system. Here, \(f = f(\boldsymbol{x},\boldsymbol{\xi},t)\) is the particle distribution function (also called the density distribution function) for particles traveling with a velocity \(\boldsymbol{\xi}\) at time \(t\) and position \(\boldsymbol{x}\). \(\boldsymbol{\Omega}\) is the particle collision operator, typically truncated to two-body collisions. The formulation given for $\boldsymbol{\Omega}$ in eqn.~\ref{eqn:BTE} is the simplification proposed by Bhatnagar-Gross-Krook (the so-called BGK collision operator) to make the BTE solvable for near-equilibrium non-trivial flows. In this operator, $\tau$ is the relaxation time and $f^{eq}$ is the Maxwellian equilibrium distribution function 
\begin{equation*}
    f^{eq} = \frac{\rho}{(2 \pi RT)^{d/2}} \exp\left( -\frac{|\boldsymbol{\xi} - \boldsymbol{u}|^2}{2RT}\right),
\end{equation*}
with density $\rho$, gas constant $R$, temperature $T$, number of spatial dimensions $d$, and bulk fluid velocity $\boldsymbol{u}$. The macroscopic variables for isothermal flows are then defined by 
\begin{align}
    \nonumber \rho &= \int f d\boldsymbol{\xi},\\
    \rho \boldsymbol{u} &= \int f \boldsymbol{\xi} d\boldsymbol{\xi}.  \label{eqn:IntMacro}
\end{align}
%
%
\subsubsection{Discretization of the velocity space}
The discretization of the velocity space, i.e. $\boldsymbol{\xi} \rightarrow \cup_{i=0}^{N} \boldsymbol{e}_i$, dictates the accuracy of the DBE as a reduced order model of the 7-dimensional BTE. For this purpose, typically Gauss-Hermite (GH) quadrature is used in conjunction with the moment expansion method proposed by Shan et al. \cite{Shan1997} and He et al. \cite{He1997APriori}  The procedure is briefly described below.

The distribution function is first expanded using the orthonormal Hermite polynomials in $\boldsymbol{\xi}$ associated with the rank-\emph{m} tensor $\mathcal{H}^{(m)}$ and weight function $\omega$ as follows
\begin{align}
    \label{eqn:Herm1}
    f(\boldsymbol{x},\boldsymbol{\xi},t) = \omega(\boldsymbol{\xi})\mathlarger{\mathlarger{ \sum}_{m=0}^\infty}\frac{1}{m!}\boldsymbol{a}_{\boldsymbol{i}}^{(m)}(\boldsymbol{x},t)\mathcal{H}_{\boldsymbol{i}}^{(m)}(\boldsymbol{\xi}),
\end{align}
where
\begin{align*}
    \boldsymbol{a}_{\boldsymbol{i}}^{(m)}(\boldsymbol{x},t) = \int f(\boldsymbol{x},\boldsymbol{\xi},t) \mathcal{H}_{\boldsymbol{i}}^{(m)}(\boldsymbol{\xi}) \, d\boldsymbol{\xi}.
\end{align*}
Here, the index $\boldsymbol{i}$ refers to the $m$-fold indices $i_1i_2...i_m$. The summation in eqn.~\ref{eqn:Herm1} can be truncated to $M$th order while retaining the first $M$ moments due to the orthonormality of the Hermite polynomials such that
\begin{align}
    \label{eqn:Herm3a}
     f(\boldsymbol{x},\boldsymbol{\xi},t) \approx f^{M}(\boldsymbol{x},\boldsymbol{\xi},t) = \omega(\boldsymbol{\xi})\mathlarger{\mathlarger{ \sum}_{m=0}^{M}}\frac{1}{m!}\boldsymbol{a}_{\boldsymbol{i}}^{(m)}(\boldsymbol{x},t)\mathcal{H}_{\boldsymbol{i}}^{(m)}(\boldsymbol{\xi}).
\end{align}
The first few Hermite polynomials are 
\begin{align*}
    \mathcal{H}^{(0)}(\boldsymbol{\xi}) &= 1  \\
    \mathcal{H}^{(1)}_{i_1}(\boldsymbol{\xi}) &= \xi_{i_1}\\
    \mathcal{H}^{(2)}_{i_1i_2}(\boldsymbol{\xi}) &= \xi_{i_1} \xi_{i_2} -\delta_{i_1i_2},
\end{align*}
yielding
\begin{align*}
    \boldsymbol{a}^{(0)} &= \rho,\\
    \boldsymbol{a}^{(1)} &= \rho \boldsymbol{u},\\
    \vdots
\end{align*}
We can discretize the velocity and expand $f^{eq}$ up to 2nd order as follows  
\begin{align}
    \label{eqn:feq}
    f^{eq}_i &= \omega_i \rho \left[1+\frac{\boldsymbol{e}_i\cdot\boldsymbol{u}}{RT} + \frac{\left(\boldsymbol{e}_i\cdot\boldsymbol{u}\right)^2}{2(RT)^2} - \frac{\boldsymbol{u}^2}{2RT} \right],
\end{align}
where $\boldsymbol{e}_i$ is the discretization of velocity $\boldsymbol{\xi}$ in the $i^{th}$ direction and  $\omega_i$ is the corresponding Gauss-Hermite weight.

This helps simplify the BTE and reduce it to 3+1 dimensions, allowing us to obtain the DBE with BGK collision operator
\begin{equation}
    \label{eqn:DBE}
    \frac{\partial f_i}{\partial t} + \boldsymbol{e}_i\cdot\nabla f_i =  -\frac{1}{\tau}\left(f_i - f^{eq}_i \right).
\end{equation}
The above equation is valid only for low Mach number isothermal flows due to the functional form of $f^{eq}$ in eqn.~\ref{eqn:feq}, with errors of the order $\mathcal{O}(Ma^2)$. While this functional form is sufficient for the flows considered in this article, higher-order expansions including the dependency on temperature can be incorporated to improve its accuracy.\cite{Shan2006}

The macroscopic flow variables in eqn.~\ref{eqn:IntMacro} can be obtained from the discrete distribution functions for $N$-point discretization of $\boldsymbol{\xi}$, as follows
\begin{align*}
    \rho &= \Sigma_i f_i,\\
    \rho\boldsymbol{u} &= \Sigma_i \boldsymbol{e}_i f_i.
\end{align*}

\subsubsection{Choice of discrete velocities.}
In two-dimensional flows, the D2Q9 velocity set is obtained from the tensor product of 1D 3-point GH quadrature rules of order 5. Here, the discrete velocities $\boldsymbol{e}_i$ are given by
\begin{align*}
    e_x &= \sqrt{3}\{0,1,0,-1,0,1,-1,-1,1\},\\
    e_y &= \sqrt{3}\{0,0,1,0,-1,1,1,-1,-1\},
\end{align*}
with corresponding weights 
\begin{align*}
    w &= (1/36)\{16,4,4,4,4,1,1,1,1\},
\end{align*}
and with lattice speed of sound $c_s=1$. This quadrature set allows the DBE formulation to capture hydrodynamics up to the same fidelity as the Navier-Stokes equations. Higher-order quadrature can be used to obtain hydrodynamics beyond the Navier-Stokes limit. Table \ref{tab:quadratures} shows a few higher-order GH quadratures in one dimension.\cite{Kim2008} The corresponding 2D quadratures can be obtained as tensor products of the 1D quadratures.  
\begin{table*}[t]
\caption{One-dimensional quadratures of various orders\label{tab:quadratures}}
\centering
\begin{tabular}{|p{3cm}|p{1.4cm}||p{5.6cm}|p{5.6cm}|}
 \hline
 Quadrature & Order & velocities & weights\\
 \hline\hline
 D1Q4   &   7    & $\pm \sqrt{3-\sqrt{6}} $& $(3+\sqrt{6})/12$ \\
 \ & \ & $\pm \sqrt{3+\sqrt{6}}$ & $(3-\sqrt{6})/12$ \\
 D1Q5   &   9    &  $0$ &  $8/15$ \\
 \  & \  &  $\pm \sqrt{5-\sqrt{10}}$ & $(7+2\sqrt{10})/60$ \\
 \  & \  &  $\pm \sqrt{5+\sqrt{10}}$ & $(7-2\sqrt{10})/60$ \\
 D1Q6  &   11   &  $\pm 0.616706590193136$ & $4.088284695558080 \times 10^{-1}$ \\
 \ & \ & $\pm 1.88917587775414$ & $8.861574604199542 \times 10^{-2}$ \\
 \ & \ & $\pm 3.32425743355142$ & $2.555784402056898 \times 10^{-3}$ \\
\hline
\end{tabular}
\end{table*}

While high-order quadratures are usually beneficial in capturing the non-linearity in the solution distribution in high Knudsen number flows, GH quadrature provides depreciating benefits due to the quadrature points far from the centroid having very small weights.~\cite{Guo2013} To alleviate this issue, Newton-Cotes (NC) quadrature can be used to obtain the discrete velocities. 
%
%
\subsubsection{Relaxation time and influence of Knudsen number} 
The dependency on Kn is introduced in the above formulation via the relaxation time $\tau$. Upon defining the mean free path $\lambda = \sqrt{3}\tau c_s$ and recalling that $\text{Kn} = \lambda/L$, we get
\begin{align}
    \label{eqn:KnTau}
    \tau = \text{Kn} L/\sqrt{3}c_s.
\end{align}
This is consistent with the standard definition of $\text{Kn} = \sqrt{3/2}\alpha^{-1}$ with $\alpha = \frac{L}{\tau}\sqrt{2k_BT}$, where $k_B$ is the Boltzmann constant. Note: this is the usual definition of Kn for Direct Simulation Monte Carlo (DSMC) methods.\cite{Ansumali2007} In addition, eqn.~\ref{eqn:KnTau} can be reparameterized depending on the expression chosen for $\lambda$. Kim and Pitsch \cite{Kim2008} use $\lambda$ derived from a first principles understanding of Kinetic Theory and arrive at an additional factor of $\sqrt{\pi/6}$ in the expression for Kn. Further corrections can be introduced by scaling $\tau$ with a Kn-dependent function such as $\psi(\text{Kn}) = \frac{2}{\pi}\arctan(\sqrt{2} \text{Kn}^{-3/4})$.\cite{Zhang2014} 

To enable comparisons to the work presented by Ansumali et al,\cite{Ansumali2007} we shall use the definition in eqn.~\ref{eqn:KnTau} in the remainder of this article. 

%
%
\subsubsection{Boundary conditions for wall-fluid interactions}
%
The most common wall boundary condition in the LBM, the so-called bounce-back scheme and its variants, is designed to enforce the no-slip boundary condition.\cite{ladd1994numerical,sukop2006dt,Krueger2003} The absence of specular reflection (i.e., bounce-forward) prevents the capturing of relative motion between the fluid and the wall leading to a slip velocity.\cite{Krueger2003} Conversely, the bounce-forward-only scheme reproduces elastic collisions with an ideal wall by enforcing angle of reflection to equal angle of incidence of the incoming distribution function at a boundary node. This scheme enables uninhibited slip at the wall. Based on these ideas, a variety of methods have been proposed to accurately capture the wall-fluid interaction in the slip-flow and transition regimes. \cite{Wang2016a}

%

Scattering-type boundary conditions enable a convenient middle-ground between uninhibited-slip and no-slip boundary conditions. To account for the non-elasticity of collisions and a rough wall, the diffuse-scattering (also known as the Maxwellian diffuse-reflection) boundary condition is typically used. Here, it is assumed that the fluid particles undergoing collision with the wall scatter back following the Maxwell distribution function, losing the memory associated with their movement prior to the collision. While in LBM, this can be implemented in different ways \cite{Sofonea2005}, we shall follow the method proposed in references \cite{Ansumali2007,Kim2008,Shi2011} as stated below. 

For the distribution function $f_i$ satisfying the criteria $(\boldsymbol{e}_i - \boldsymbol{u_b}) \cdot \boldsymbol{n} < 0$, where $\boldsymbol{u_b}$ is the true wall velocity and $\boldsymbol{n}$ is the \emph{outward} pointing normal, the diffuse-scattering kernel yields \\
\begin{align}
        \label{eqn:FullBC}
        f_i(\boldsymbol{x_b},t) = \frac{\Sigma_{(\boldsymbol{e}_j-\boldsymbol{u_b})>0} \lvert(\boldsymbol{e}_j-\boldsymbol{u_b}) \cdot \boldsymbol{n} \rvert  f_j}{\Sigma_{(\boldsymbol{e}_k-\boldsymbol{u_b})<0} \lvert(\boldsymbol{e}_k-\boldsymbol{u_b}) \cdot \boldsymbol{n} \rvert f_k^{eq}} f_i^{eq}(\rho_b,\boldsymbol{u}_b),
\end{align}
which reduces to
\begin{align}
        \label{eqn:UsedBC}
        f_i(\boldsymbol{x_b},t) = f_i^{eq}(\rho_b,\boldsymbol{u_b}),
\end{align}
for steady unidirectional flows.\cite{Kim2008} 

Depending on the physics required to be described in the problem, further improvements to the diffuse-scattering kernel can be incorporated. The Maxwell-type second-order slip model estimates the slip velocity as a function of Kn, first- and second-derivatives of fluid velocity normal to the wall, and parameterized slip coefficients.\cite{Wang2016a} The values of the coefficients are obtained by other models, such as a micro-scale molecular dynamics simulation.\cite{Chibbaro2008} The Langmuir slip boundary condition resolves this issue by estimating the slip velocity as a function of the wall velocity, fluid velocity adjacent to the wall, and physical coefficients of the gas particles for a given interaction potential.\cite{eu1987nonlinear,abe1997d} Nevertheless, in this article we shall only consider the simplified diffuse-scattering boundary condition described in eqn.~\ref{eqn:UsedBC}, as it has been shown to be in good agreement with the DSMC solutions of the BTE for the flows considered here.\cite{Ansumali2007,Guo2013}

%
%
\subsection{Discontinuous Galerkin Finite Element Method}

%
%
\subsubsection{Weak formulation of DG-FEM}
Consider the spatial domain D tessellated by non-overlapping, \emph{d}-dimensional, elements $\text{D}^k$ of characteristic size \emph{h}, forming the mesh $\mathcal{T}_h$. While the size \emph{h} can vary among the elements, it is required that the faces of the elements along the perimeter of the mesh conform exactly to the domain. Unlike continuous Galerkin FEM, we duplicate values of variables at nodal points $\boldsymbol{x}^k$ on the vertices, edges, and faces of the elements in order to ensure locality of the scheme within each element. We denote the boundary of each element  with $\partial \text{D}^k$ and associate with it an outward-pointing normal $\boldsymbol{n}$. The global solution $f_i$ can be approximated by a piecewise $p$-th order polynomial over $\mathcal{T}_h$ as follows.

On each element, the local solution is approximated by a polynomial basis $\eta(\boldsymbol{x})$ as
\begin{align*}
    \boldsymbol{x}\in\text{D}^k : f_{i,h}^k(\boldsymbol{x},t) = \sum_{j=1}^{N_p} f_{i,h}^k(\boldsymbol{x}_j^k,t)\eta_j^k(\boldsymbol{x}).
\end{align*}
Here, $f_{i,h}^k(\boldsymbol{x}_j^k,t)$ is the nodal value of the approximate solution $f_{i,h}^k$ at one of the $N_p$ nodal points. $\eta$ can be chosen to be the interpolating Lagrange polynomial defined in 1D as $l_m = \prod_{m=1,m \neq n}^{p+1} \left( \frac{x - x_n}{x_m - x_n}\right)$. Thereafter, we can represent the global approximate solution as 
\begin{align*}
    \boldsymbol{x}\in\text{D}: f_i(\boldsymbol{x},t) \simeq f_{i,h}(\boldsymbol{x},t) = \sum_{k=1}^{K} f_{i,h}^k(\boldsymbol{x},t),
\end{align*}
where $K$ is the total number of elements. Consider setting $\boldsymbol{F}_i = \boldsymbol{e}_i f_i$, where one should note that the two-fold appearance of the index $i$ does \emph{not} imply a summation over $i$. With this in mind, the local residual on element $\text{D}^k$ can be written as 
\begin{align*}
    \mathcal{R}_{i,h}^{k} (\boldsymbol{x},t) = \frac{\partial f_{i,h}^k}{\partial t} + \boldsymbol{\nabla} \cdot  \boldsymbol{F}_{i,h}^{k} + \frac{1}{\tau}(f_{i,h}^{k} - f_{i,h}^{eq,k}), 
\end{align*}
in accordance with eqn.~\ref{eqn:DBE}.
Upon multiplying the above expression with a test function and integrating over the domain $\text{D}^k$, we impose the condition
\begin{equation*}
    \int_{\text{D}^k} \phi_j^k(\boldsymbol{x}) \mathcal{R}_{i,h}^{k}(\boldsymbol{x,t}) \, d\boldsymbol{x} = 0. 
\end{equation*}
Here, $\phi_j^{k} (\boldsymbol{x})$ is an arbitrary test function, typically chosen to be a Lagrange polynomial. Upon integrating by parts and substituting the expression for residual $\mathcal{R}_{i,h}^{k}$, we get
\begin{align}
    &\int_{\text{D}^k} \phi_{j}^{k} \frac{\partial f_{i,h}^k}{\partial t} d\boldsymbol{x} - \int_{\text{D}^k} \boldsymbol{F}_{i,h}^k \cdot \boldsymbol{\nabla}\phi_{j}^{k} d\boldsymbol{x} 
    +\frac{1}{\tau} \int_{\text{D}^k} (f_{i,h}^k - f_{i,h}^{eq,k})\phi_{j}^{k} d\boldsymbol{x} = - \int_{\partial {\text{D}^k}} \phi_{j}^{k} \boldsymbol{F}_{i,h}^* \cdot \boldsymbol{n} \,  ds. \label{eqn:byparts} 
\end{align}
In the expression above, $\boldsymbol{F}^*$ is the numerical flux which allows information to pass between neighboring elements. Two elements sharing a (\emph{d}-1)-dimensional face \emph{F} are considered to be face-neighbors. We denote the normal vector pointing from the positive to the negative side of a shared face on element $\text{D}^k$ with $\boldsymbol{n}^+$; and correspondingly $\boldsymbol{n}^-$ is the normal vector pointing in the opposite direction. At this point, we introduce the notation for the mean and jump operators as
\begin{align*}
        \{\{&f_i\}\} = \frac{f_i^- + f_i^{+}}{2}, \\
        [[&f_i]] = \boldsymbol{n}^-f_i^- +  \boldsymbol{n}^+f_i^+.
\end{align*}
The numerical flux $\boldsymbol{F}^*_i$ can be represented as
\begin{align*}
    \boldsymbol{F}^*_i = (\boldsymbol{e}_if_i)^* = \{\{\boldsymbol{e}_if_i\}\}+\lvert \boldsymbol{e}_i \rvert \beta[[f_i]],
\end{align*}
where $\beta$ is a parameter that corresponds to the numerical dissipation added to the scheme. $\beta = 0.5$ leads to the central-flux, inducing no additional dissipation, and $\beta = 1$ reproduces the upwind flux.

%
\subsubsection{Implementation of boundary conditions}
The boundary condition described in eqn.~\ref{eqn:UsedBC} is a Dirichlet boundary condition that is imposed over the boundary in the DG method by setting the value exterior to the element equal to a prescribed value, e.g.~$f_i^+ = f_i(\rho_b,\boldsymbol{u_b})$. For the elements on the boundary, this yields 
\begin{align*}
        \{\{&f_i\}\} = \frac{f_i^- + f_i^{eq}(\rho_b,\boldsymbol{u_b})}{2}, \\
        [[&f_i]] = \boldsymbol{n}^-f_i^- +  \boldsymbol{n}^+f_i^{eq}(\rho_b,\boldsymbol{u_b}).
\end{align*}
The same approach is followed for the full form of the BC described in eqn.~\ref{eqn:FullBC}.

\subsubsection{Time marching}

To maintain stability and accuracy, we used Singly Diagonally Implicit Runge-Kutta (SDIRK)~\cite{alexander1977diagonally,burrage1982efficiently} methods to discretize eqn.~\ref{eqn:byparts}. Unlike explicit RK schemes, these implicit schemes allow a time step that is not constrained by the CFL limit. 

A four-stage 4\textsuperscript{th}-order scheme was chosen for the order of accuracy analysis on micro-Couette flow (see the next section), and a 1-stage first-order RK method, which reduces to the backward difference method, was chosen for the more demanding micro-cavity flow simulations. The overall implementation was performed within the Solution Adaptive Numerical Solver (SANS).~\cite{galbraith2015verification,galbraith2018sans} Newton's method was used to linearize the non-linear system at each stage of the SDIRK methods, and each linear system was subsequently solved using the Generalized Minimal Residual Method (GMRES).\cite{saad1986gmres} The iterative solution was obtained using the Portable, Extensible Toolkit for Scientific Computation (PETSc). \cite{balay2017petsc,galbraith2018sans}


%
%
\section{\label{sec:results}Results}
\subsection{Micro-Couette flow}
\begin{figure}[h!]
    \centering
    \includegraphics[scale=0.35]{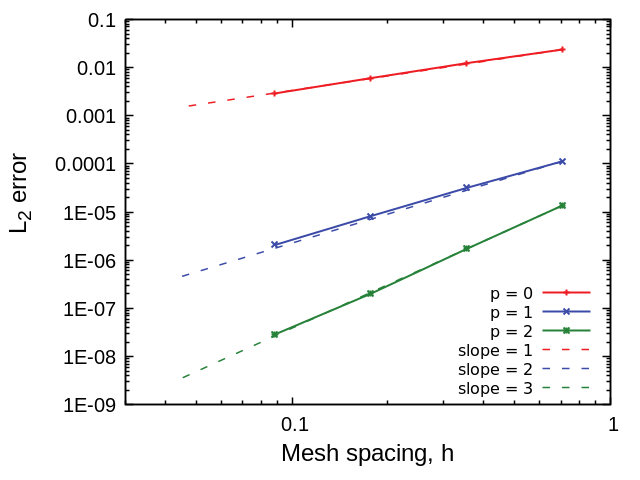}
    \caption{$L_2$ error in the streamwise velocity $u_x$ for Couette flow vs. mesh size $h$ for different polynomial degrees $p$.}
    \label{fig:OOA}
\end{figure}

To evaluate the accuracy of the scheme, we first consider Couette flow for fluids with 0 < Kn $\leq$ 1.5. Here, we consider two parallel plates separated by a distance \emph{L} confining the fluid in the \emph{y}-axis. The top and bottom walls are prescribed with opposing velocities of magnitude $u_{w,x} = 0.16 c_s$. The plates are placed at \emph{y} = 1 and \emph{y} = 0 respectively. The wall BC is imposed using eqn.~\ref{eqn:UsedBC} while periodic boundary conditions are applied to the domain boundaries along the \emph{x}-axis. Following the definitions set by Asumali et al.\cite{Ansumali2007}, the analytical solution to \emph{x}-velocity from the linearized DBE for the D2Q16 velocity set is given by
\begin{align}
    \label{eqn:exact}
    u_x = \frac{1}{Z_{16}}\sinh \left(\frac{y+\frac{1}{2}}{\text{Kn}L}\right)\Delta U + \frac{1}{\Theta_{16}}\left(\frac{y+\frac{1}{2}}{L}\right)\Delta U+U, 
\end{align}
where
\begin{align*}
    \Delta U &= U_{top} - U_{bottom}, \\
    U &= (U_{top} + U_{bottom})/2, \\
    \mu &= \sqrt{3-\sqrt{6}} + \sqrt{3+\sqrt{6}}, \\
    \Theta_{16} &= 1 + 2\text{Kn}\left[\frac{2\cosh\left(\frac{1}{2\text{Kn}}\right)+\mu \sinh\left(\frac{1}{2\text{Kn}}\right)}{\mu \cosh\left(\frac{1}{2\text{Kn}}\right)+2\sqrt{3}\sinh\left(\frac{1}{2\text{Kn}}\right)}\right],\\
    Z_{16} &= \frac{\mu}{4\text{Kn}}\Bigg[(4\text{Kn}+\mu)\cosh\left(\frac{1}{2\text{Kn}}\right) 
    2(\mu \text{Kn}+\sqrt{3})\sinh\left(\frac{1}{2\text{Kn}}\right)\Bigg].
\end{align*}
To evaluate the order of accuracy of our DG-FEM, we consider the D2Q16 velocity set  at Kn = 1 for a variety of uniform structured grids with 8, 32, 128, and 512 triangle elements using polynomial interpolation degrees $p = 0, 1, \text{and}$ 2. Here, the structured grids were generated by splitting Cartesian grids into triangles, e.g. the 32 element grid was formed by splitting a $4\times4$ Cartesian grid.
%
The $L_2$-error is calculated using
\begin{align*}
    E^{L_2} = \sqrt{ \sum_{k=1}^K \int_{\text{D}^k} (u_x^e(\boldsymbol{x}) - u_x^k(\boldsymbol{x}))^2 d{\boldsymbol{x}}},
\end{align*}
where $u_x^e$ is the exact solution evaluated using eqn.~\ref{eqn:exact} and $u_x^k$ is the solution obtained from the DG-FEM. The error is evaluated after $t = 40 $ characteristic time units to ensure steadiness in the solution. In each case, we confirmed that $\lvert (\rho,u_x,u_y)_{40} - (\rho,u_x,u_y)_{30}\rvert < 10^{-12}$. A time step of $dt = 0.05$ was chosen for time marching. 
Figure \ref{fig:OOA} shows the $L_2$ error for different polynomial orders and mesh sizes. The slope of each curve provides the order of accuracy for a given polynomial order. We find that the calculated order of accuracy is very close to the theoretically expected value of $p+1$.

\begin{figure*}
    \centering
    \includegraphics[scale=0.275]{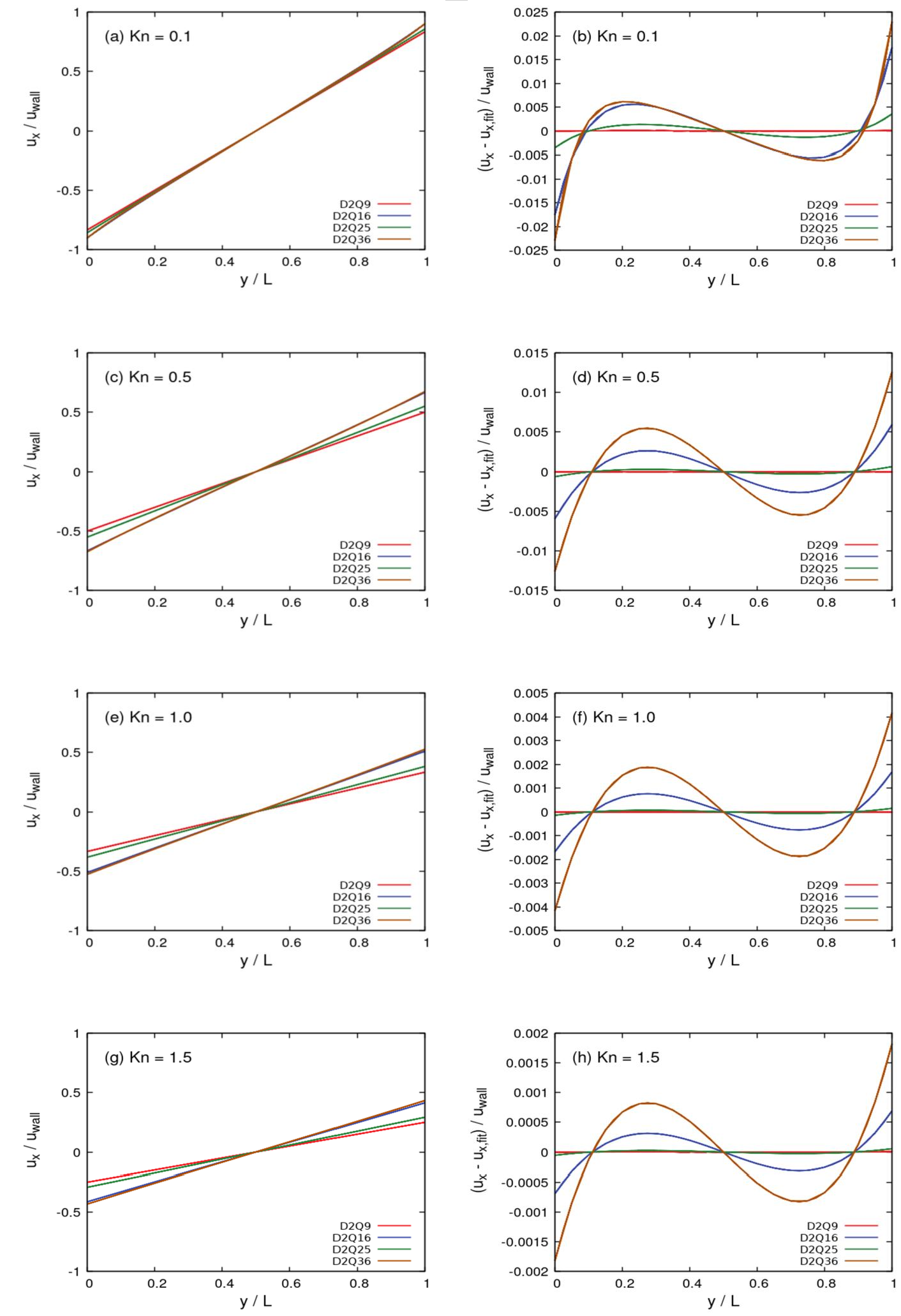}
    \caption{Velocity profiles for Couette flow with different velocity sets and Knudsen numbers. (Left) Normalized streamwise velocity profiles. (Right) Predicted Knudsen layer defined as a deviation from a straight line profile constrained to pass through \{$\tfrac{1}{2}$, 0\}. }
    \label{fig:profile}
\end{figure*}
\begin{figure*}
    \centering
    \includegraphics[scale=0.35]{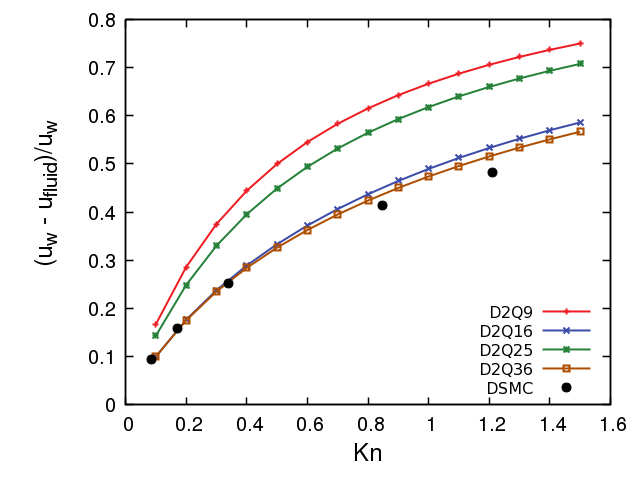}
    \caption{Comparison of normalized slip velocity for Couette flow at different values of Kn. DSMC values replotted using data from Ansumali et al.\cite{Ansumali2007}}
    \label{fig:slip}
\end{figure*}

Following the order of accuracy study, we performed a set of numerical experiments with different Knudsen numbers and velocity sets, using a 400-element structured grid with $p=2$.
The left column in Figure~\ref{fig:profile} shows the normalized streamwise velocity profile for different values of Kn. For all values of Kn, we find that the profiles corresponding to odd velocity sets D2Q9 and D2Q25 are close to each other, and the profiles corresponding to even velocity sets D2Q16 and D2Q36 are near to each other in a similar fashion. As expected, D2Q9 underpredicts the slope of the velocity the most, in addition to overestimating the slip effects at the walls. We note that D2Q16 and D2Q36 are closer to the expected results, whereas D2Q25 predicts the incorrect slope and slip velocity although it is obtained using a quadrature rule of higher order than D2Q16.~\cite{Kim2008,Shi2011} This is reflected in the prediction of the so-called Knudsen layer, as shown in the right column of Figure~\ref{fig:profile}. The Knudsen layer, a characteristic of a flow that violates the continuum assumption, is the deviation of the velocity profile from a straight line as predicted by the Navier-Stokes equations. Here, the Navier-Stokes velocity profile ($u_{x,fit}$) is obtained using a least-squares fit of a straight line constrained to pass through the point $\{\frac{1}{2},0\}$ of the velocity profile ($u_x$). The limitations of D2Q9 are explicitly demonstrated by the complete absence of any deviation from the straight line for all Kn. While D2Q25 predicts a non-negligible extent of Knudsen layer formation, it underpredicts the effect substantially in comparison to D2Q16 and D2Q36.  At $\text{Kn} = 0.1$, we find a good match between D2Q16 and D2Q36 results. However, at $\text{Kn} \geq 0.5$ there is a departure of the flow-field predicted by D2Q16 from D2Q36. This attribute is well known, as D2Q16 begins to deviate from the correct flow-field (from DSMC or the linearized Boltzmann equation) at $\text{Kn} \simeq 0.5$.\cite{Meng2011} 

Figure \ref{fig:slip} shows the normalized slip velocity evaluated at the wall for different velocity sets and values of Kn, in conjunction with DSMC results for reference purposes. As expected, D2Q9 and D2Q25 overpredict the slip velocity, whereas, D2Q16 and D2Q36 maintain a trend close to the DSMC results. While D2Q36 delays the onset of the deviation from the correct slip velocity profile in comparison to D2Q16, we find that beyond $\text{Kn} \simeq 0.8$, velocity sets with higher accuracy are required. It must be noted that even at $\text{Kn} = 0.1$, we find a slight mismatch in the Knudsen layer formation predictions of the D2Q16 and D2Q36 models. This indicates the importance of using higher-order quadrature rules in the discretization of the velocity in the phase-space. However, as mentioned previously, D2Q25 underperforms in comparison to the lower-order D2Q16 quadrature. This is explained by Shi et al.~\cite{Shi2011} as a boundary condition artifact due to the mis-alignment of the discrete velocities in odd-numbered velocity sets. The discrete particle velocities parallel to the boundaries, i.e., parallel to the x-axis in the system simulated, do not contribute to the half-space moments required to be described at the boundaries. 
Hence, the velocity sets with particle velocities parallel to the wall boundaries introduce errors of larger magnitude, especially when they are associated with larger weights. Therefore, we would expect the flow-field using D2Q49 to be closer to D2Q25 than D2Q36, 
and higher-order even-numbered schemes such as D2Q64 are more likely to match the DSMC results up to $\text{Kn} = 1$ (see Meng and Zhang~\cite{Meng2011}).

\subsection{Micro-Cavity flow\label{sec:microcavity}}

\begin{figure*}
    \centering
    \includegraphics[scale=0.25]{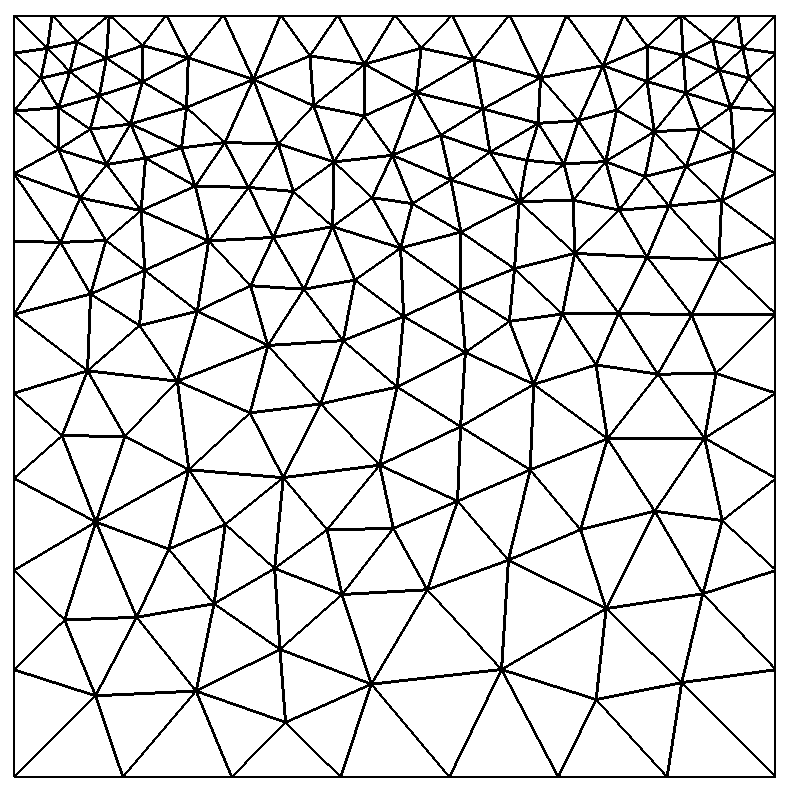}
    \caption{Unstructured mesh comprised of 315 triangle elements used in the micro-cavity simulations.}
    \label{fig:mesh}
\end{figure*}

We also applied the DG-FEM to a lid-driven cavity flow with Knudsen numbers of 1, 2 and 8. Here, a domain of characteristic length $L$ = 1 was chosen with the Maxwellian diffuse-scattering boundary condition (eqn.~\ref{eqn:FullBC}) applied on all four boundaries. 
The lid at $y = L$ is prescribed a velocity corresponding to a Mach number of 0.16. The simulations were performed on a uniform structured mesh comprised of 200 triangle elements of order $p$ = 1. For comparison, simulations at $\text{Kn}=1$ were also performed on a uniform structured mesh with 450 triangle elements and an unstructured mesh with 315 triangle elements. The unstructured mesh is shown in Figure~\ref{fig:mesh}. The solution was deemed steady if the flow-fields $(u_x,u_y)$ evaluated 5 seconds apart varied by $\mathcal{O}(10^{-5})$ or less. Velocity discretizations based on Gauss-Hermite and Newton-Cotes quadrature were used. In particular, D2Q16, D2Q64, D2Q144, D2Q289 and D2Q324 velocity sets were considered based on the tensor products of 1D Gauss-Hermite quadrature of order 7, 15, 23, 33 and 35 respectively. In addition, the tensor product of 1D Newton-Cotes quadrature with 32 intervals (33 nodes, order 32), spanning the domain [-4$c_s$, 4$c_s$] $\times$ [-4$c_s$, 4$c_s$], was considered to obtain a NC D2Q1089 velocity set.

\begin{figure*}
    \centering
    \includegraphics[scale=0.3]{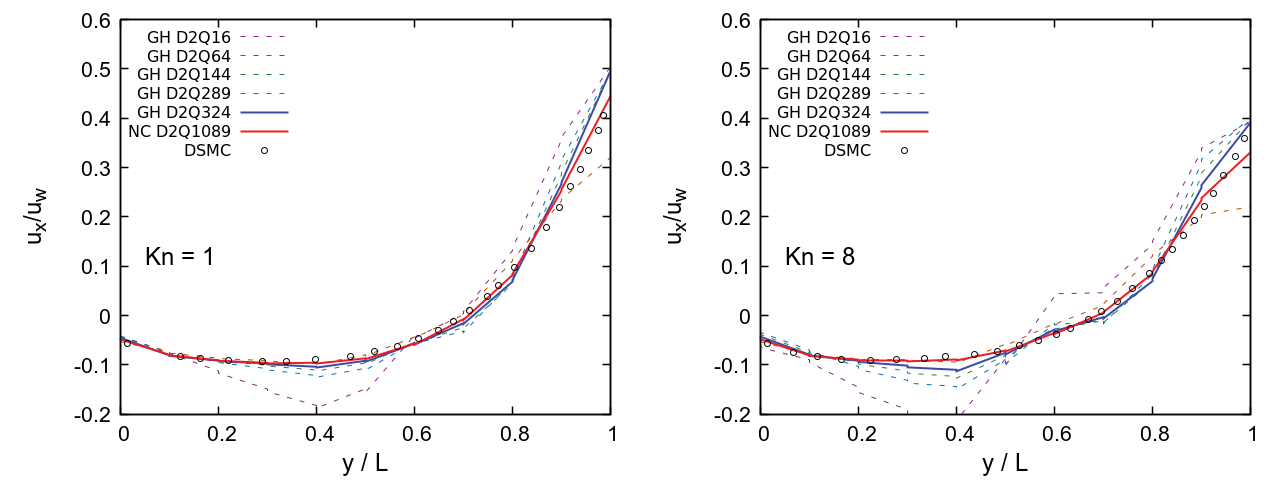}
    \caption{Comparison of $u_x$ along the cavity center line for different GH and NC quadratures for (left) $\text{Kn}=1$, (right) $\text{Kn}=8$. The DSMC data is replotted using results from reference \cite{Guo2013}.}
    \label{fig:LDquad}
\end{figure*}

Figure \ref{fig:LDquad} shows the $u_x$ profile along the micro-cavity center \{x = 0.5\} at $\text{Kn}=1$ and $\text{Kn}=8$ for different velocity sets, in conjunction with DSMC results for reference purposes. The DG-FEM results were all obtained on the 200-element grid. We note that D2Q16 exhibits large discontinuities in the first-derivative of the profile for $\text{Kn}=1$, although it was adequate for the Couette flow cases presented in the previous section. These discontinuities are due to the \emph{ray-effect}, i.e.~the preferential alignment of the macroscopic velocities along the directions of the discrete velocity space. Higher-order quadratures -- D2Q64, D2Q144, D2Q289, D2Q324, and D2Q1089 show improvements in matching the DSMC results. One should note that D2Q289 overestimates the slip velocity at the top wall due to the nature of the BC definition, as was seen for odd quadrature rules in Couette flow. However, other high-order quadrature rules, NC D2Q1089 and GH D2Q324  match the DSMC results well. At $\text{Kn} = 8$, we find some slight discontinuities in the velocity profile with GH D2Q324, but NC D2Q1089 maintains a relatively smooth profile. As expected, GH D2Q16--289 predict sharp discontinuities. Here, the fact that NC D2Q1089 has more velocity directions seems to give it an advantage over GH D2Q324, despite its basis on a quadrature rule of slightly lower strength. We will offer an explanation for this phenomenon in what follows.

The sharp gradients introduced by the walls cause highly irregular regions of the flow to form locally, and hence low-order quadrature schemes, such as D2Q16, are insufficient. To reduce the unphysical oscillations caused by the deviation of the solution from the local equilibrium, higher numbers of discrete velocities are required. However, as we mentioned previously, GH-quadrature rules provide depreciating benefits with increasing  quadrature strength because the weights of the points far away from the centroid have vanishingly small values. For example, the D2Q324 model obtained from a tensor product of the D1Q18 quadrature rule has 24 points with weights lower than machine zero ($2.2 \times 10^{-16}$), with the lowest weight of order $10^{-23}$. Hence, these 24 points do not contribute to the evaluation of the solution. Now, the NC D2Q1089 velocity set also contains points that do not contribute to the solution. However, only 4 points have weights below machine-zero, unlike 24 in GH D2Q324. In addition, D2Q1089 has significantly more velocity directions overall. Broadly speaking, the accuracy of NC D2Q1089 (despite having an odd number of points), emphasizes the importance of requiring many, more equally weighted velocity directions relative to GH velocity sets. 

\begin{figure*}
    \centering
    \includegraphics[scale=0.3]{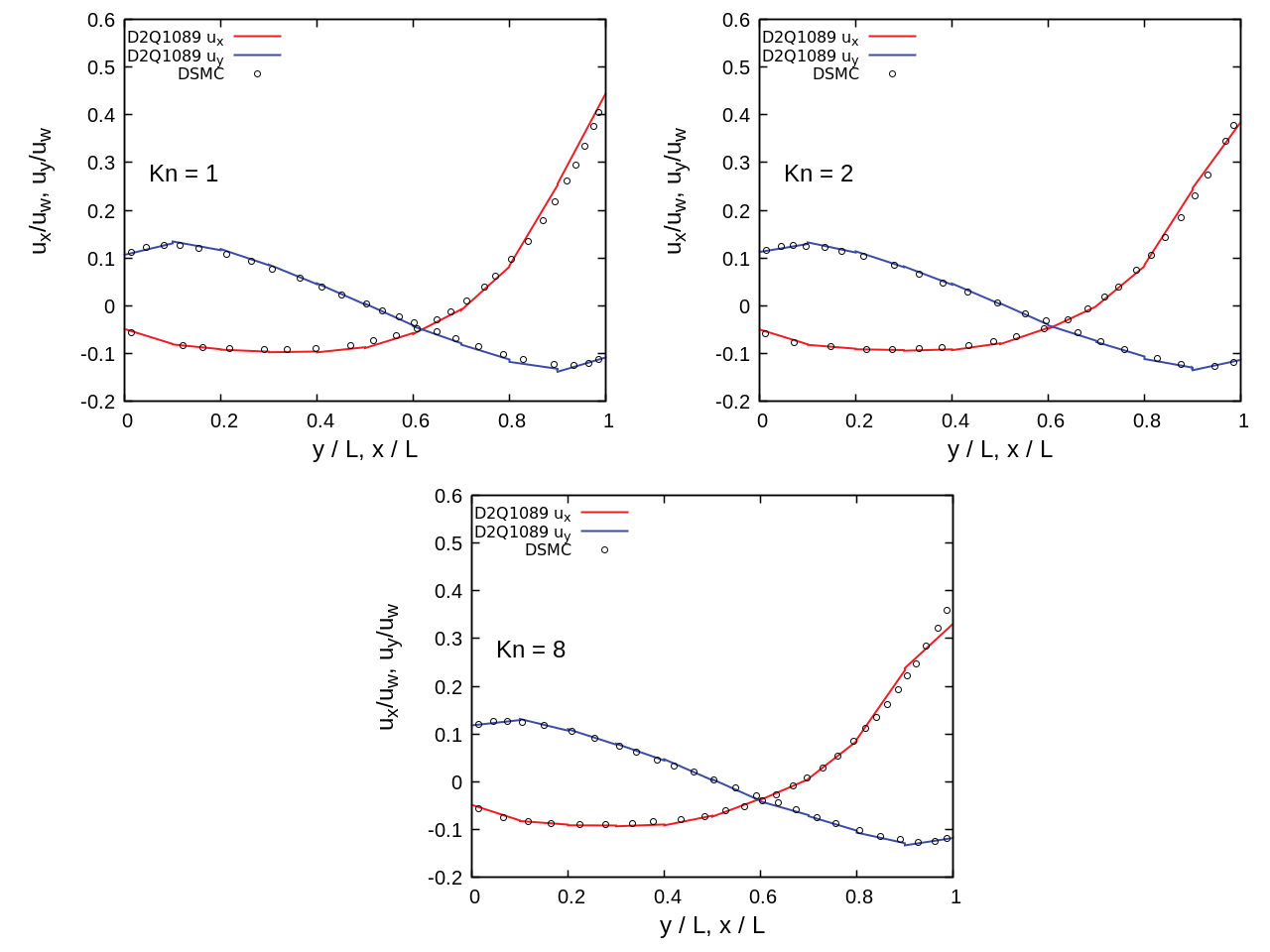}
    \caption{Normalized $u_x$ (red) and $u_y$ (blue) along the cavity center for different Knudsen numbers. The DSMC data is replotted using results from reference \cite{Guo2013}. }
    \label{fig:LDprofiles}
\end{figure*}

\begin{figure*}
    \centering
    \includegraphics[scale=0.1575]{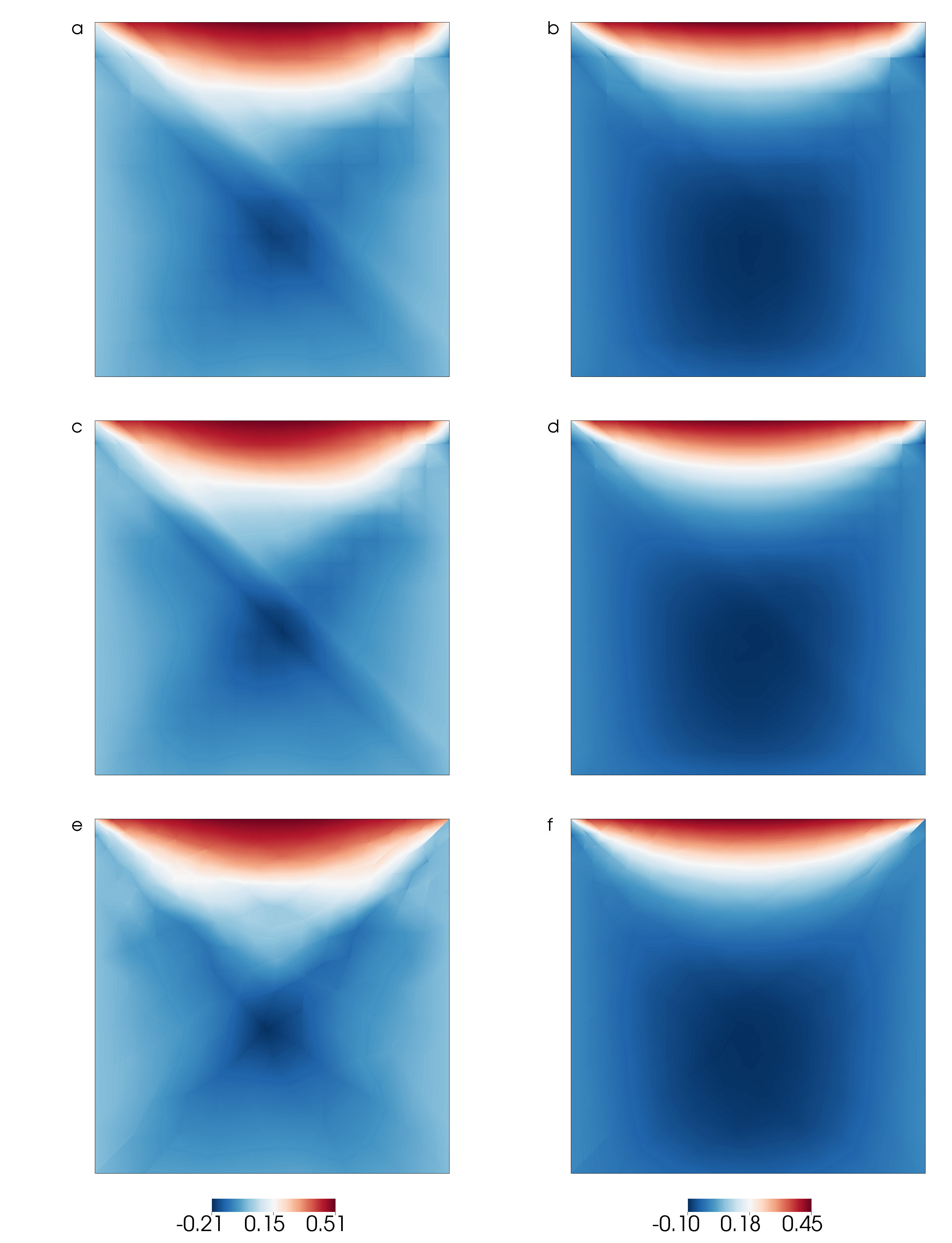}
    \caption{Contours of normalized $u_x$ at Kn = 1 for (left) GH D2Q16 and (right) NC D2Q1089. (a-b) uniform structured mesh with 200 triangle elements, (c-d) uniform structured mesh with 450 triangle elements, and (e-f) unstructured mesh with 315 triangle elements.}
    \label{fig:LDcontour}
\end{figure*}

Lastly, one may consider some additional results for the NC D2Q1089 velocity set. Figure~\ref{fig:LDprofiles} shows $u_x$ and $u_y$ profiles at all values of Kn, obtained on the 200-element grid, in conjunction with DSMC results. Furthermore, Figure \ref{fig:LDcontour} shows contour plots that illustrate the significant reduction in the ray-effect with NC D2Q1089 compared against GH D2Q16 at $\text{Kn} =1$.  Even for the NC velocity set, we find some slight evidence of the ray-effect's presence along the corners next to the moving wall in Figure \ref{fig:LDcontour}b for the 200-element uniform structured grid. Figures \ref{fig:LDcontour}c-f show the contours for the 450-element uniform structured and 315-element unstructured meshes, highlighting that the ray-effect is (mostly) grid-independent. The slight evidence of the effect at the upper corners suggests that NC D2Q1089 may not completely capture the significant non-linearity, unlike at the cavity center where it matches the DSMC results well. Guo et al.~\cite{Guo2013} suggest using the NC D2Q10000 velocity set to capture this non-linearity. Nonetheless, our results are reasonable, while requiring far fewer velocity directions. Yet, we acknowledge that the number of velocity directions will need to be significantly increased for $\text{Kn} \gg 1$.



\section{Conclusion} \label{sec:conclude}
We have introduced a DG-FEM for solving the discrete Boltzmann BGK equation, for the purpose of capturing non-continuum effects in isothermal fluid flows. We show that the scheme exhibits a spatial convergence of order $p+1$ with a D2Q16 velocity set, for a Couette flow at $\text{Kn} = 1$. We note that high-order accuracy is achieved with a local stencil, unlike finite volume and finite difference methods. In addition, the temporal accuracy is not limited to second order as the full coupling between the streaming and collision parts of the DBE is retained. Therefore, different time integration schemes, such as backward Euler (first order) or Runge-Kutta (any order), can be used depending on the temporal accuracy needed for one's application. Although only implicit schemes were implemented in this work to ensure stability for large time-steps, the retention of the coupling allows the use of explicit time integration methods of any order as well.

In addition, we analyzed the slip velocity and the Knudsen layer predicted in Couette flow for different velocity sets (D2Q9, D2Q16, D2Q25, and D2Q36). For the Maxwellian diffuse-scattering boundary condition, we found that the even-numbered quadratures were more accurate as all the discrete velocities contributed to the boundary condition, unlike odd-numbered quadratures. However, larger velocity sets were required for higher Knudsen numbers, as we found the onset of departure from the DSMC results at $\text{Kn} \simeq 0.5$ for D2Q16 and $\text{Kn} \simeq 0.8$ for~D2Q36. 

The deficiency of low-order quadrature was exacerbated in a more complex, lid-driven micro-cavity flow. For $\text{Kn} = 1$, we compared GH velocity sets D2Q16, D2Q64, D2Q144, D2Q289 and D2Q324, with NC D2Q1089. We found that most GH velocity sets show significant first-derivative discontinuities in the flow-field due to the ray-effect. More specifically, while NC D2Q1089 (order 32) exhibited a smooth profile and matched well with DSMC data at all Knudsen numbers, GH D2Q324 (order 35) showed the characteristic discontinuities at $\text{Kn} = 8$. This is because high-order GH quadrature has points clustered near the boundaries with weights close to machine-zero, diminishing the contributions of its velocity directions. Conversely, NC quadrature contains points uniformly distributed with most weights larger than machine-zero, enabling the scheme to better capture the non-linearity of the flow-field. However, we still found the slight presence of the ray-effect with NC D2Q1089 near the upper corners of the micro-cavity, indicating the need to use larger velocity sets to accurately capture the flow-field as the Knudsen number increases. 

\section*{Data availability}
The data that support the findings of this study are available from the corresponding author upon reasonable request.

\clearpage

\bibliography{AllCollection}

\end{document}